\documentclass[letterpaper,8pt]{extarticle}  
\usepackage{import}
\usepackage{local}
\usepackage[left=.65in,top=.65in,bottom=.65in,right=.65in]{geometry}
\usepackage{lipsum}
\usepackage{natbib}
\usepackage{amsmath}
\usepackage{amssymb}
\usepackage{gensymb}

\title{Mid-IR hyperspectral imaging with undetected photons}

\author{%
\textbf{Marlon Placke\textcolor{Accent}{\textsuperscript{1,*}}, %
Chiara Lindner\textcolor{Accent}{\textsuperscript{2}}, %
Felix Mann\textcolor{Accent}{\textsuperscript{1}}, %
Inna Kviatkovsky\textcolor{Accent}{\textsuperscript{1}},
Helen M. Chrzanowski\textcolor{Accent}{\textsuperscript{1}}, %
Hendrik Bartolomaeus\textcolor{Accent}{\textsuperscript{3}}, %
Frank K\"{u}hnemann\textcolor{Accent}{\textsuperscript{2,4}}, %
Sven Ramelow\textcolor{Accent}{\textsuperscript{1,5,*}}
}\\
\begin{small}\textcolor{Accent}{\textsuperscript{1}}Institute for Physics, Humboldt Universität zu Berlin, Newtonstra\ss e 15, 12489 Berlin, Germany\\ 
\textcolor{Accent}{\textsuperscript{2}}Fraunhofer Institute for Physical Measurement Techniques IPM, Georges-K\"{o}hler-Allee 301, 79110 Freiburg, Germany\\
\textcolor{Accent}{\textsuperscript{3}}Institute of Experimental Biomedicine, University Hospital Würzburg, Josef-Schneider-Stra\ss e 2, 97080 Würzburg, Germany\\
\textcolor{Accent}{\textsuperscript{4}}Institute of Physics, Universit\"{a}t Freiburg, Hermann-Herder-Stra\ss e 3, 79104 Freiburg, Germany\\
\textcolor{Accent}{\textsuperscript{5}}Ferdinand-Braun-Institut, Gustav-Kirchhoff-Stra\ss e 4, 12489 Berlin, Germany\\
\textcolor{Accent}{\textsuperscript{*}}Correspondence: \textcolor{Accent}{mplacke@physik.hu-berlin.de, sven.ramelow@physik.hu-berlin.de} \\ \end{small}
}

\date{}
\begin{document}
\maketitle
\thispagestyle{empty}

\section{Abstract}

\noindent
\textbf{\textcolor{Accent}{Sensing with undetected photons has become a vibrant, application-driven research domain with a special focus on the mid-infrared (mid-IR) wavelength region. Since the mid-IR contains spectral bands with highly specific and strong molecular absorbance signatures, often referred to as fingerprints, a multitude of different samples and their compositions can be detected and quantified spectroscopically. Enhancing this inherently sample alteration-free spectroscopic method with imaging capabilities leads to a powerful technique for environmental monitoring and biomedical applications that enables automated diagnostics while omitting time-consuming and non-reversible labeling steps.
To evade the shortcomings of state-of-the-art instruments for mid-IR hyperspectral microscopy related to cost, complexity, power-consumption, and performance, which are associated with technological challenges for mid-IR cameras and low-noise and broadband mid-IR sources, here, we construct a proof-of-concept nonlinear interferometer in a wide-field imaging arrangement combined with high-resolution spectral acquisition by pixelwise quantum Fourier transform infrared spectroscopy.  For the broadband range of 2300-3100 cm$^{-1}$, covering the important CH-stretch band, we perform hyperspectral imaging that simultaneously resolves 3500 spatial modes, each with a spectral resolution of  10 cm$^{-1}$ leveraging in total around  10$^5$ spatio-spectrally entangled photon modes. Our image acquisition uses a commercial, megapixel sCMOS camera, while a medium-power and compact c.w. pump laser is the only necessary light source. For a moderate speed of 360 voxel/s yielding a dominantly shot-noise influenced signal-to-noise ratio (SNR) of 50, we demonstrate the practicality of our novel hyperspectral imaging technique for microplastics detection and bio-imaging tasks, and outline engineering solutions to increase its speed by several orders of magnitude. This shows that our quantum imaging technique is highly promising for applications requiring compact, cost-effective label-free analyses.}}
\begin{doublespacing}
\section{Introduction}
Vibrational spectroscopy addresses the molecular resonances of optically transmissive samples to reveal their composition-specific fingerprints in a label-free and non-invasive manner.  Consequently, the method has become indispensable in numerous analytics tasks ranging from bio-medical and forensic screening, to gas-sensing, food and drug-monitoring, and quality control of synthetic and industrially engineered materials.
However, since these quantized excitations are largely situated in the mid-infrared (mid-IR) range of the electromagnetic spectrum, straightforward measurements are limited by the performance of mid-infrared illumination sources as well as their corresponding sensor technology.

The interest to specifically investigate the spatio-spectral structures of complex samples e.g. to differentiate healthy from malignant biological tissue, furthermore motivates hyperspectral imaging methods. Here, the three dimensional data sets (i.e. hyperspectral cube of wavenumber-specific spectral images or position-specific spectra) are acquired and reconstructed in broadband excitation imaging systems. Among the various considerable approaches range methods with both mid-IR light generation (e.g. using fast-scanning quantum cascade lasers or frequency combs) and detection with single pixel mid-IR (mercury cadmium telluride) detectors in combination with a sample scanning-based imaging system \cite{Reihanisaransari2024-yv, 10.1063/5.0225616, GONZALEZCABRERA201874}. 
Other methods sidestep the comparatively noisy and low-resolution mid-IR detection, but still rely on specialized illumination sources e.g. broadband upconversion imaging from supercontinuum generation~\cite{Fang2024-fj}. Finally, Raman spectroscopy imaging allows for both (near-)visible excitation and detection, but requires high intensity illumination due to inherently small scattering cross sections \cite{ozeki2020molecular}. All of these techniques yield specific tradeoffs with regard to spectral bandwidth, acquisition speed, setup complexity and component costs, imaging quality, and noise equivalent powers (NEP).

Another alternative pathway to decouple the sensing wavelength from that of the probe is to exploit quantum sensing with undetected photons. Here, the mid-IR probe light can be generated from inherently low-noise nonlinear conversion processes. Given a suitable nonlinear crystal medium, broadband mid-IR idler photons may be generated alongside their near-IR signal photon counterparts from the spontaneous conversion of its sum frequency pump. Since the incorporation of this conversion process into an interferometric arrangement enables the transfer of the idler-probed sample information onto the signal light, the detection may then be realized with near-IR sensors, i.e. circumventing the relative performance penalty of mid-IR detectors and allowing access to the high-efficiency, low-noise, multi-pixel detection CCD and CMOS technologies. While various implementations of such quantum sensing schemes were focused on either spatial or spectral analysis \cite{Vanselow:20, kviatkovsky2020microscopy, lindner2021nonlinear}, we combine both degrees of freedom to access the full three-dimensional cube of mid-IR hyperspectral information using only visible and near-IR pump sources and detection.
This enables a novel hyperspectral imaging approach with both high spectral and spatial resolution in a low-complexity, low-noise, and cost-effective setup. After characterizing the experimental setup and finding a signal-to-noise ratio near the shot noise limit, we demonstrate its suitability for analyzing complex samples,  
such as microplastics annotation and label-free histology.
\section{Experimental Setup}

\begin{figure}[!htbp]
    \centering
\includegraphics[width=11cm]{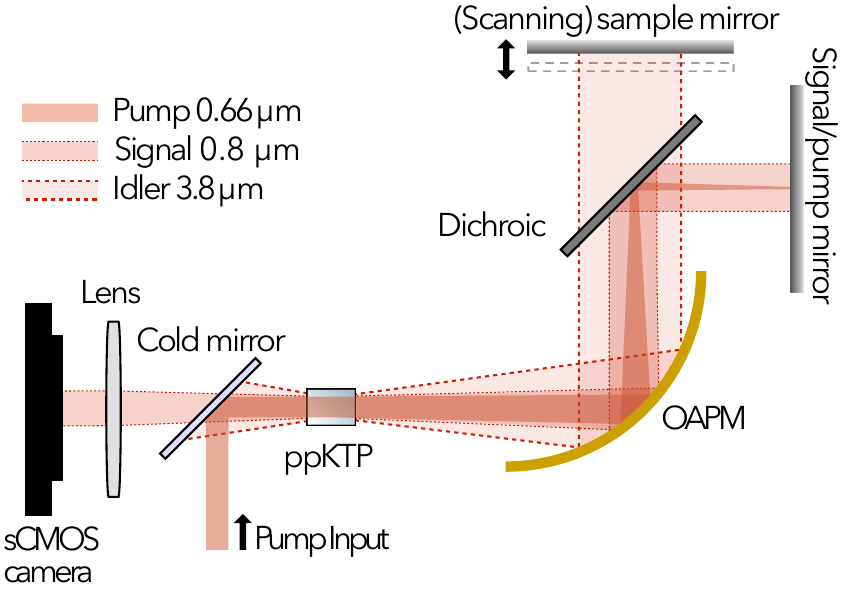}
   \caption{\small{{\bf Experimental setup:} For the highly spatially multimodal hyperspectral acquisition, a large-aperture ppKTP crystal is pumped by a 660 nm continuous wave laser, collimated to a Gaussian waist of 1.1 mm. The resulting highly non-degenerate signal and idler fields are collimated by an off-axis parabolic mirror and reflected onto a folded-Mach-Zehnder-style arrangement of dichroic and retro-reflecting silver mirrors. Upon a second nonlinear crystal passing of the pump and overlapping of first and second pass-generated signal and idler fields, the near-infrared signal is imaged onto a scientific CMOS camera for detection. The hyperspectral information cube of pixel- (wavenumber-) specific spectra (images) is obtained from a Fourier transform analysis of the pixelwise interferograms recorded upon full-range scans of the idler mirror stage.}}
    \label{fig:setup}
\end{figure}

For sensing with undetected photons, we use a nonlinear interferometer in SU(1,1) configuration (Fig. \ref{fig:setup}). The pump source is a single-frequency continuous wave diode pumped solid-state laser with 310 mW of pump power at $\lambda_{p0} =$ 659.7 nm, which is collimated to a Gaussian waist of $w_0 \approx$ 1.1 mm at the center of the nonlinear crystal. As a source of spectrally and spatially entangled photons, we use a periodically-poled potassium titanyl phosphate (ppKTP) crystal with a large aperture of 4x8 mm and a length of $L_{\textrm{c}} =$ 3 mm. The crystal's periodic poling of $\Lambda =$ 20.2 \textmu m is chosen to satisfy phase-matching for highly non-degenerate, collinear signal-idler emission bands centered around 0.8 and 3.8 \textmu m, respectively. This combination of wavelengths and crystal medium exploits a group velocity matching condition~\cite{Vanselow:19} such that the signal-idler emission becomes particularly broadband, with more than 20 THz of combined bandwidth.  Notably, the large crystal aperture allows for a large pump waist, which, by imparting only a small transverse momentum uncertainty to the resulting signal and idler photon pairs gives rise to a highly spatially multimode nonlinear emission with more than 3500 modes enabling a wide-field imaging functionality with a reasonably large FOV.

To minimize chromatic aberrations, the generated signal and idler fields are collimated and reflected by an off-axis parabolic mirror (OAPM) placed at a focal length equal distance of $f_{\textrm{oapm}} =$ 152 mm from the crystal center. Next, a dichroic mirror reflects the converging pump and collimated signal onto a fixed-position mirror, while the collimated idler is transmitted onto a second mirror with samples placed on it in transflective probing. The sample mirror is mounted on a piezo stage for precise scanning of the interferometer arm length. The two interferometric mirrors are positioned at a distance of a focal length from the OAPM such that pump, signal, and idler are reverted back onto their initial propagation by the OAPM. 

As pump, signal, and idler pass the nonlinear crystal a second time, interference occurs between the first and second signal-idler generation processes and the sample's spatio-spectral transmission properties sampled by the idler from the first generation process are imprinted on the interference observable in the signal light, while the idler light can be discarded. 
The back-propagating signal light is then collimated by a convex lens placed at a focal length's distance from the crystal center ($f_1$), which effectively relays the sample plane onto a 4 megapixel backside-illuminated sCMOS camera for detection. This arrangement produces an image magnification of $M \approx n_{s0} \lambda_{i0} f_{\textrm{oapm}}/(n_{i0} \lambda_{s0} f_{1}$) as determined by the focal lengths of the used focusing elements. An additional lens may be placed in the detection path to adapt the magnification factor independently, trading off single pixel SNR against a suitable oversampling rate to preserve the spatial information given by the optical resolution of the imaging system.

\section{Results}
\subsection{Imaging performance}

\begin{figure}[!htbp]
\centering
\includegraphics[width=15cm]{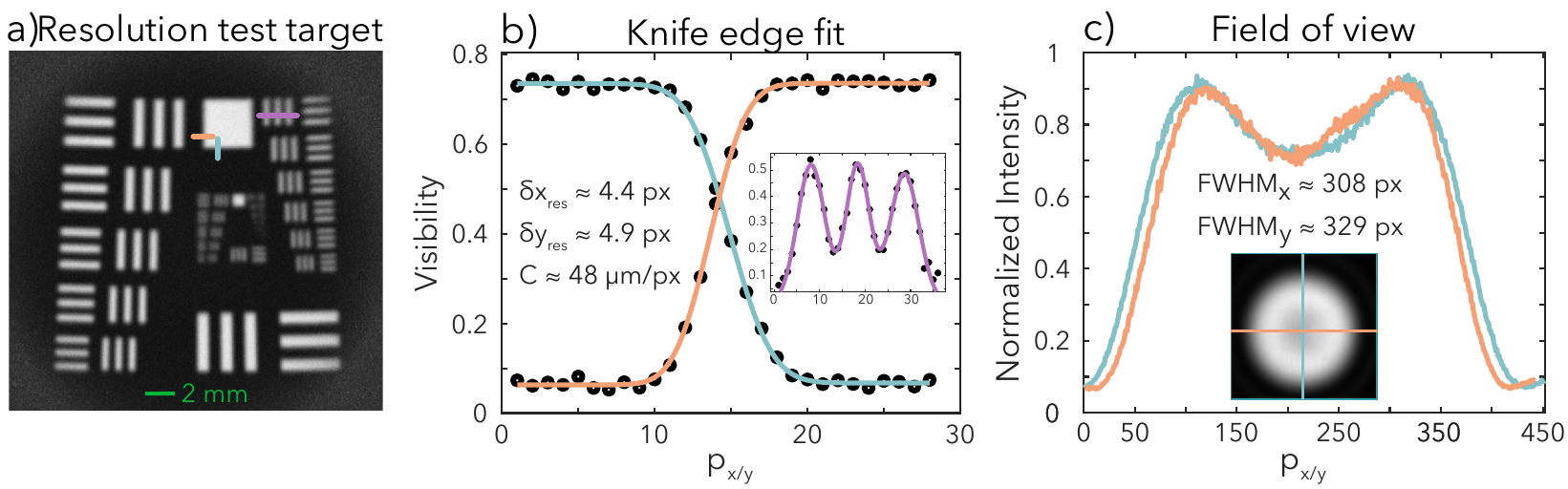}
\caption{\small{{\bf Image resolvability characterization:} a) The resolution target visibility image was obtained as the normalized pixelwise maximum contrast upon scanning the idler arm shortly about the zero-path length difference point. b) Line profiles along the square and bar structures of the test target served to determine the pixel-to-length calibration $C$ and yielded a spatial resolution of $\delta x_{\textrm{res}} \approx$ 210 $\pm$ 16 \textmu m and  $\delta y_{\textrm{res}} \approx$ 236 $\pm$ 21 \textmu m, respectively. c) Line profiles of the pixelwise sum of maximum and minimum counts (inset) yielded the field of view as $FOV_{\textrm{x/y}} \approx$ 14.8 / 15.8 $\pm$ 0.2 mm amounting to a number of spatial modes $N_x \approx$ 70 $\pm$ 6 and $N_y \approx$ 67 $\pm$ 7 in orthogonal directions of the sample plane.}}
\label{fig:Vis_Rtest}
\end{figure}

To experimentally characterize the imaging properties, we placed a clear optical path 1951 USAF resolution test target in close vicinity to the idler mirror surface and scanned the interferometer over a distance of $\Delta L_{\textrm{test}} =$ 3.6 \textmu m (corresponding to 2.1 periods of the nonlinear interference) around the zero path length difference position. To reduce the effect of chromatic artifacts on the raw visibility images, we filtered the signal light with a narrow dielectric bandpass corresponding to an idler transmission range of $\Delta \tilde{\nu}_{\textrm{i}} \approx$ 2980-3020 cm$^{-1}$. 
For the acquisition of the interferogram, the stage was continuously scanned with a speed of $\dot{r}_{\textrm{scan}} = $ 120 nm/s and images were captured with an exposure time of $t_{\textrm{exp}} =$ 1 s. Generally, a trade-off was made between the step distance ($\Delta L_{\textrm{sd}} = \dot{r}_{\textrm{scan}} \cdot t_{\textrm{exp}}$ = 120 nm) of consecutive captures and the total acquisition time ($t_{\textrm{tot}} = \Delta L_{\textrm{max}} \cdot t_{\textrm{exp}}/\Delta L_{\textrm{sd}}$).
The idler-probed (quantum) image was then retrieved as the pixel-wise maximum normalized contrast in the number of counts in the set of captures as 
\begin{equation}
    \textrm{VIS}(p_x, p_y) = \frac{N_{\textrm{max}}(p_x, p_y) - N_{\textrm{min}}(p_x, p_y)}{N_{\textrm{max}}(p_x, p_y) + N_{\textrm{min}}(p_x, p_y)} \qquad . \label{eq:vis}
\end{equation}
The visibility image of the resolution target is shown in Figure \ref{fig:Vis_Rtest}a.
From the group one, element one bar structures of the test target (purple line in Fig. \ref{fig:Vis_Rtest} a \& b) we found a pixel-to-length calibration of $C =$ 1 mm /(20.8 $\pm$ 0.2 px) between sample features and their camera-recorded image. 
Using the adjacent square structure for knife edge-like line profiles and fitting an error function to the data in Fig. \ref{fig:Vis_Rtest}b, we obtained a spatial resolution of $\delta x_{\textrm{res}} \approx$ (4.4 $\pm$ 0.3 \ px) $\cdot \ C  \approx$ 210 $\pm$ 16 \textmu m and  $\delta y_{\textrm{res}} \approx$ (4.9 $\pm$ 0.4  px) $\cdot \ C  \approx$ 236 $\pm$ 21 \textmu m (full width half maximum (FWHM)) of the Gaussian derivative of the error function.
Here, the separate x- and y-values represented a trade-off between the resolution in each transversal direction.
This trade-off stemmed from a slight astigmatism to the images given the small magnification. For larger magnifications (corresponding to $C \approx$ 1 mm/ 35 px), astigmatism was negligible, suggesting a significant contribution from insufficient oversampling of the sharp-edged test target structures.

Finally, the FOV was estimated from line profiles along the vertical and horizontal direction of the illumination cone, see Figure~\ref{fig:Vis_Rtest}c. These yielded a $FOV_{\textrm{x/y}} \approx$ 14.8 / 15.8 $\pm$ 0.2 mm at FWHM corresponding to a number of spatial modes of $N_x = FOV_{\textrm{x}} / \delta x_{\textrm{res}} \approx$ 70 $\pm$ 6 and $N_y = FOV_{\textrm{y}} / \delta y_{\textrm{res}} \approx$ 67 $\pm$ 7. The experimentally obtained imaging performance agreed well with the model prediction of 80 modes for a resolution limit of 181 \textmu m and a FOV of 14.5 mm that followed from the JSPI calculations. Apart from aberrations, imperfect alignment, and, most notably, the insufficient spatial oversampling given the choice of magnification, the finite filter bandwidth contributed to the minor discrepancy between the experimentally observed and theoretically anticipated image resolvability.

\subsection{Spectral acquisition characterization}

\begin{figure}[!htbp] 
\centering
\includegraphics[width=15cm]{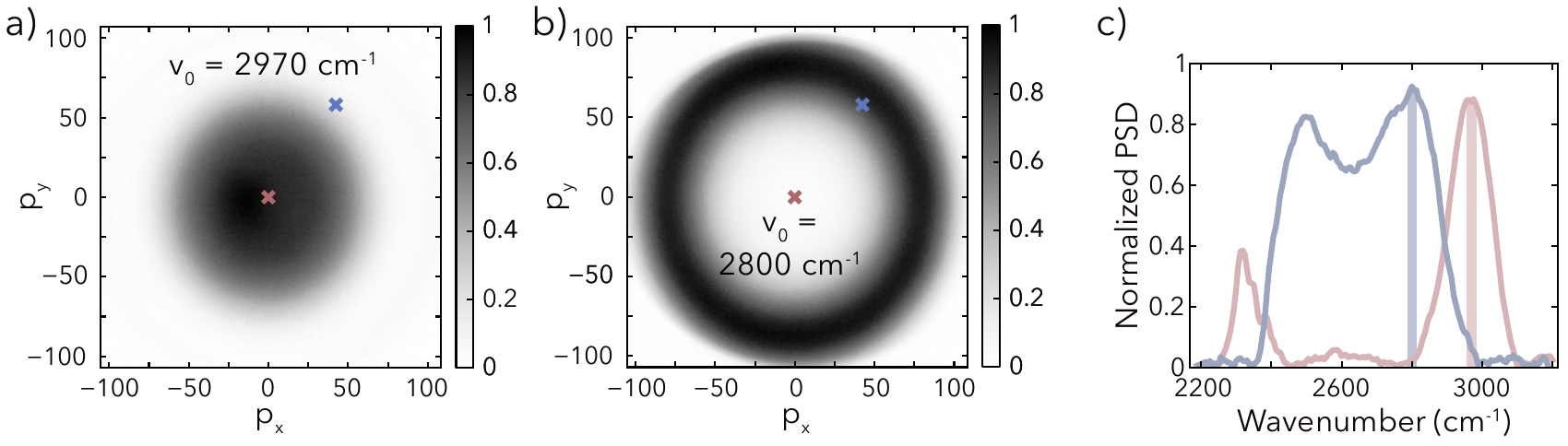} 
\caption{\small{{\bf Spatio-spectral structure of the nonlinear emission:} a, b) Normalized wavenumber-specific illumination of the sample plane at 2970 cm$^{-1}$ and at 2800 cm$^{-1}$, respectively. c) Position-specific spectra for the x-marked pixels in the two spectral images. For the collinear emission, a dual-band idler spectrum was found, which extended to CO$_2$ absorption features below 2400 cm$^{-1}$. For non-collinearly positioned pixels, the dual-peak structure merged around intermediate wavenumbers. A rough estimate of the spectral bandwidth of $\Delta \nu_i \approx$ 400 $\pm$ 50 cm$^{-1}$} was made by eye measure since the dual peak structure as well as the CO$_2$ absorption features imposed large ambiguity on the FWHM quantification. The spatio-spectral illumination structure can be eliminated through adaption of the poling period and/or the crystal temperature, see Supplemental Information.} 
\label{fig:FOVeffect}
\end{figure}

To demonstrate hyperspectral imaging and characterize the spectral acquisition, the bandpass filter was removed and the interferometer scanned over the full displacement range ($\approx \pm$ 400 \textmu m) with a constant speed while continuously acquiring camera images. The speed of the scanning stage was adapted such that within a single shot exposure time, the stage moved by $\delta L_\text{sd}=$ 120 nm, resulting in an optical path difference of 240 nm, amounting to a total acquisition time of $t_{\textrm{scan}} =$ 6667 \ $\cdot t_{\textrm{exp}}$. Since the camera images were measured with a spatial resolution of $\approx$ 4 px, a (2x2) binning was applied to each image. The binning improved the resulting pixelwise signal-to-noise ratio while largely retaining the image resolution.

To analyze the series of camera images recorded while scanning the interferometer, the intensity trace of each (binned) pixel was Fourier-transformed. We use a discrete Fourier-transform based on the Fast Fourier-transform algorithm to obtain the pixel-wise spectra. This produced the full hyperspectral information cube. Figure~\ref{fig:FOVeffect} shows the normalized Fourier-transform amplitude of the camera pixels at 2970 cm$^{-1}$ (a) and 2800 cm$^{-1}$ (b). Figure~\ref{fig:FOVeffect}c shows the normalized Fourier-transform spectra of a collinear (red) and non-collinear (blue) camera pixel (positions marked in Fig.~\ref{fig:FOVeffect}a and b).

As a consequence of the broadband phase matching with a poling period $\Lambda =$ 20.2 \textmu m and crystal temperature $T_{\textrm{exp}} =$ 87 $^{\circ}$C, the collinear emission featured dual-band signal and idler spectra (idler peaks around 2350 \& 2950 cm$^{-1}$), while the intermediate wavenumber range ($\approx$ 2600 cm$^{-1}$) was emitted under non-collinear angles. Therefore, the illumination profile varied with the choice of wavenumber (cf. Fig.~\ref{fig:FOVeffect}a, b), while the emission spectra varied with the distance from the FOV center (cf. Fig.~\ref{fig:FOVeffect}c).

\begin{figure}[!htbp]
\centering
\includegraphics[width=14.5cm]{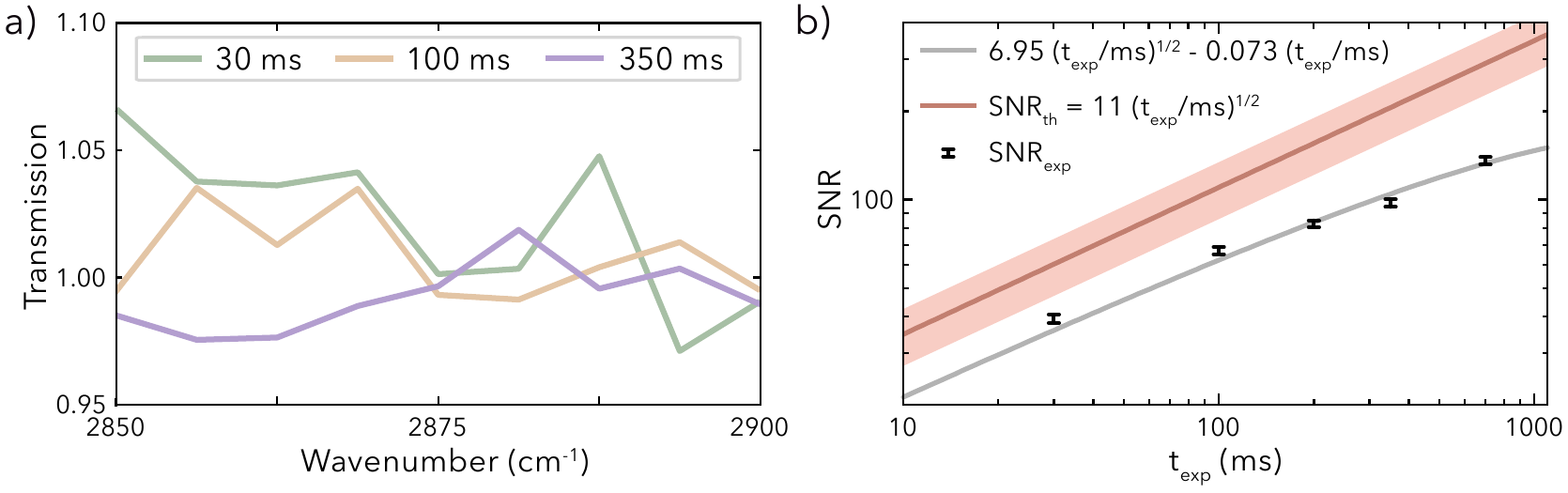} 
\caption{\small{{\bf Spectral acquisition benchmark:} a) The experimental SNR value was calculated from the ratio of the mean spectral transmission in sample-free measurements to its standard deviation between 2850 and 2900 cm$^{-1}$ for each choice of single shot exposure time. b) A nonlinear fit to the experimental SNR values revealed a linear correction term to the squareroot-proportional time dependence that likely stemmed from instabilities in the interferometric setup.  By comparison, the experimental performance already approximated the shot noise limited SNR ($SNR_{\textrm{th}}$) closely showing near-ideal utilization of the spectroscopic illumination probe.}}
\label{fig:SNRscaling}
\end{figure}

To characterize the SNR of the hyperspectral acquisition, we evaluated the sample-free transmission as the ratio of power spectral densities (PSDs) reconstructed from two consecutively recorded interferograms with identical exposure times.
The experimental SNR was calculated as the ratio of the mean value of the transmission and its standard deviation in the range of $\Delta \nu_{\textrm{snr}} =$ 2850 - 2900 cm$^{-1}$ that includes the peak PSD value as illustrated in Figure \ref{fig:SNRscaling}a. Single pixel SNR values were obtained from averages over (10x10)-pixel segments centered around positions of peak illumination strength at $\nu (N_{\textrm{max}}$).
Figure \ref{fig:SNRscaling}b gives such single-pixel SNR data derived from raw camera captures as a function of exposure time. From a nonlinear fit to the data points, we found a square root-like scaling function with a linear correction factor that likely stemmed from instabilities (e.g. mechanical vibrations or pump wavelength fluctuations) of the interferometric scan. Taking into account the fitted coefficients in
\begin{equation}
SNR_{\textrm{exp}} \approx 6.95 \cdot \sqrt\frac{t_{\textrm{exp}}}{1 \ \textrm{ms} } - 0.073 \cdot \frac{t_{\textrm{exp}}}{1 \ \textrm{ms} } \quad ,
\end{equation}
a maximum SNR around 165 was expected upon extrapolation to exposure times of about $t_{\textrm{max}} \approx$ (6.95/2/0.073)$^2$ ms $\approx$ 2.25 s, beyond which the SNR-diminishing linear term would have become dominant. Consequently, the interferometric FTIR configuration was sufficiently stable to record interferograms within the dynamic range of the camera, which saturated in camera counts for multisecond exposure. In the range of exposure times between tens and hundreds of miliseconds, the acquisition was dominated by the square-root-proportional shot noise term and a SNR of 50 was obtained for $t_{\textrm{exp}} \approx$ 62 ms. Taking into account a spectral bandwidth of $\Delta \nu_{\textrm{fwhm}} \approx$ 400 cm$^{-1}$ at a resolution of 10 cm$^{-1}$, the hyperspectral acquisition yielded $R_{\textrm{acq}} \approx$ 360 voxel per second (of total integration time) per 310 mW of pump power and a SNR of 50. We estimated the shot noise limit following~\cite{Lindner:21}
\begin{equation}
    SNR_{\textrm{th}} = \eta_{\textrm{vis}} \sqrt{\frac{n \ t_{\textrm{exp}}}{F}} \sqrt{\frac{\epsilon P_{\textrm{ppx}}}{h \nu_{s0}}} \frac{\delta \nu_{\textrm{res}}}{\Delta \nu}
\end{equation}
where the avalanche photodiode related excess noise factor $F$ was omitted for the employed low-gain CMOS camera. Further, the SPDC power was normalized to the number of illuminated pixels yielding the per-pixel power $P_{\textrm{ppx}} \approx$ 50 fW. A visibility of $\eta_{\textrm{vis}} \approx$ 65 \% and a detector efficiency of $\varepsilon(\lambda_s \approx$ 815 nm) $\approx$ 35 \% were accounted for together with a spectral resolution ($\delta \nu_{\textrm{res}} =$ 10 cm$^{-1}$) and SPDC bandwidth ($\Delta \nu \approx$ 400 cm$^{-1}$). Here, instead of multi-scan averaging as employed in fast acquisition single mode FTIR, $n$ was taken as the number of image captures to yield the total measurement time through the product with the single step exposure time $t_{\textrm{exp}}$. Comparing the calculated SNR limit of $SNR_{\textrm{th}}(t_{\textrm{exp}} =$ 100 ms) $\approx$ 110 to the corresponding experimental value of $SNR_{\textrm{exp}}(t_{\textrm{exp}} =$ 100 ms) $\approx$ 62, we found a nearly shot-noise limited performance with only small contributions of additional noise sources such as pump laser noise and interferometer instability. Nevertheless, for the sacrifice of sensitivity, i.e. when aiming for lower SNR, the hyperspectral acquisition could be accelerated quadratically, while cavity-assisted pump-enhancement~\cite{lindner2023} may still hold a ten- to hundredfold linear speed-up potential at a fixed sensitivity. 

\subsection{Polymer film identification}

\begin{figure}[!htbp]
\centering
\includegraphics[width=15cm]{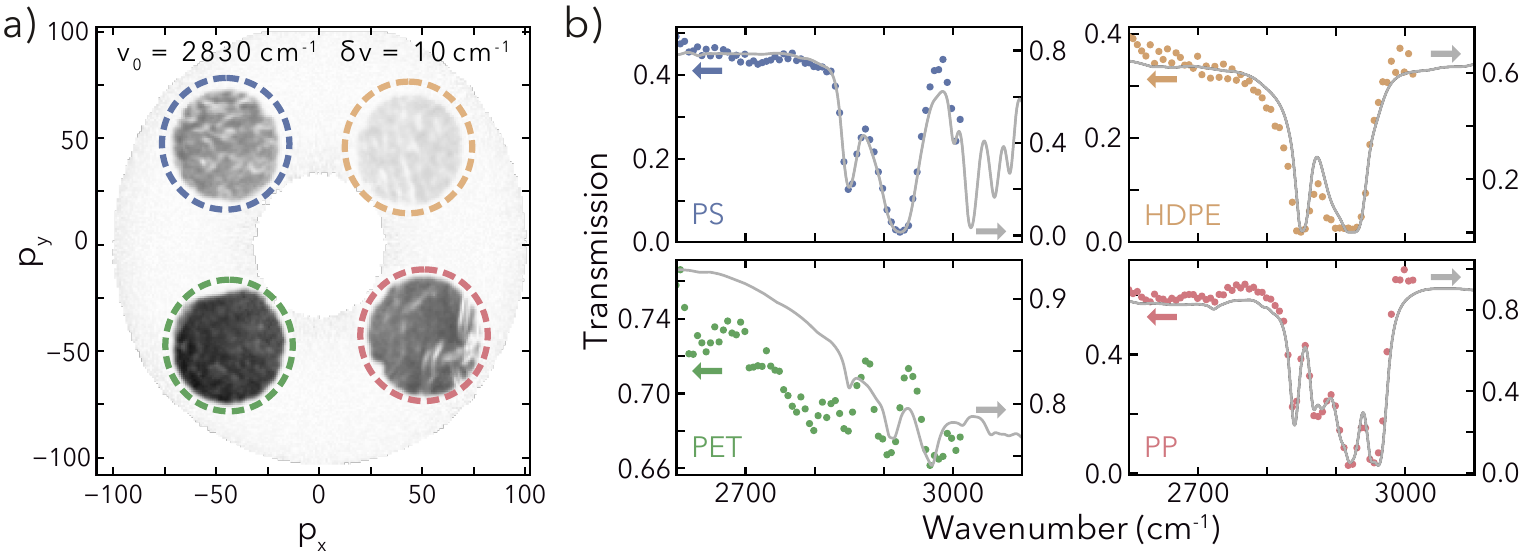} 
\caption{\small{{\bf Polymer film identification:} a) Mid-IR transmission image of the multi-polymer sample: Four bulk films (PS, HDPE, PET, PP) were placed in an aperture mask and measured with the hyperspectral imaging setup with an apodized spectral resolution of $\delta \nu_{\textrm{res}} \approx$ 10 cm$^{-1}$. b) Transmission spectra of the polymers measured with the hyperspectral imaging setup (colored dots, averaged within the circular masks shown in a)) and measured with a commercial non-imaging FTIR spectrometer (solid lines). Please note the independent scaling of the transmission axes (left: HI, right: FTIR), to account for broadband scattering losses in the former.}} 
\label{fig:multipoly}
\end{figure}

To demonstrate the spectral accuracy of the hyperspectral transmission measurements, we conducted first tests with a sample with known spatial and spectral properties.  For the measurement, bulk polymer films of polystyrene (PS), high density polyethylene (HDPE), polyethylene terephthalate (PET), and polypropylene (PP) were placed in a cardboard aperture mask with circular cutouts of 5 mm diameter (see Fig.~\ref{fig:multipoly}a). The hyperspectral images were measured with an exposure time of 200 ms and a scanning speed of 600 nm/s and the interferograms were apodized to a spectral resolution of 10 cm$^{-1}$. For each pixel, the sample transmission was calculated as the quotient of the sample spectrum and a reference spectrum. Due to the broadband non-collinear emission of the SPDC source, the spectral images received a wavelength-dependent magnification factor. The resulting spectral distortion was corrected using a rescaling operation suggested by Fang et al.~\cite{Fang2024-fj} based on a mapping of transmission features at different wavelengths.

The spectral transmissions given in Fig. \ref{fig:multipoly}b (colored dots) were calculated as multi-pixel averages over the indicated dashed circle segments.
For the reference data, the transmission of the polymer films was also measured with a commercial (non-imaging) FTIR-spectrometer ($\delta \nu_{\textrm{res}} \approx$ 4 cm$^{-1}$ spectral resolution), shown as solid lines in Fig.~\ref{fig:multipoly}b. The transmission measurement in the hyperspectral setup was more strongly influenced by scattering of the (not perfectly flat) polymer films. To account for this, the spectra are shown with independent scaling.  The measured absorption bands of all four polymers clearly coincided, even for the weak absorption band of PET. This demonstrated the suitability of the nonlinear-interferometric hyperspectral setup for polymer identification tasks.\\

\begin{figure}[!htbp]
\centering
\includegraphics[width=13cm]{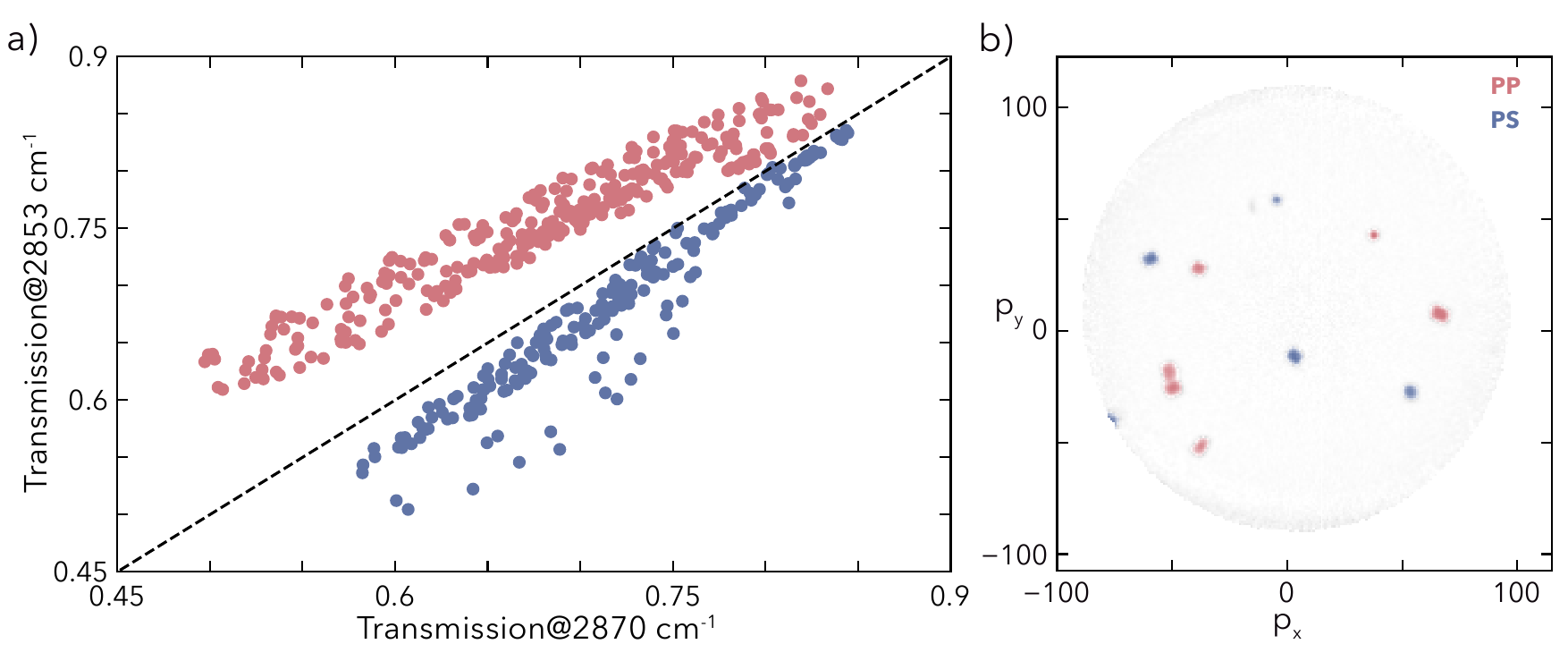} 
\caption{\small{{\bf Microplastics annotation:} a) Polymer scraps fabricated from polypropylene (PP, red) and polystyrene (PS, blue) can be distinguished based on their pixel-wise spectral transmission at 2870 cm$^{-1}$ and 2853 cm$^{-1}$. b) False color image of the classified microplastic particles based on their hyperspectral transmission demonstrated at 200 to 300 \textmu m particle size and optical resolution.}} 
\label{fig:microplastics}
\end{figure}

A central motivation of hyperspectral imaging is to identify material and tissue compositions of complex samples at a level of high spatial resolution. Here, we  performed a simplified analysis task: The classification of small polymer particles distributed within the FOV, similar to the requirements of microplastics detection. For this purpose, PS and PP films were cut to particle sizes of 200 to 300 \textmu m corresponding to the spatial resolution of the imaging arrangement. These scraps were then distributed on a carrier mirror with unknown spatial distribution. As a first step, camera pixels containing some kind of polymer were identified using the criterion that the mean transmission of the spectra at wavelengths from 2850-2870 cm$^{-1}$ fell below 80 \%. For the analysis, we assumed a sufficiently diluted sample (no overlap of the different polymer scraps) and that the spectral imprint of the scraps covered several pixels and therefore allowed for averaging. For classification, the average transmission over the neighboring 3 pixels was considered.

Figure~\ref{fig:microplastics}a shows the (multi-pixel averaged) transmission of the polymer pixels within the sample at 2870 cm$^{-1}$ and 2853 cm$^{-1}$. As described in the Supplemental Information section S2, the transmission value at these two wavenumbers can be used to effectively distinguish polystyrene (PS) and polypropylene (PP). Pixels with $T$(2853 cm$^{-1}$)<$T$(2870 cm$^{-1}$) were identified as PS (blue dots), otherwise as PP (red dots). Over a large range of transmission values (which were not only influenced by sample absorption but also other factors such as scattering and polarization effects), the classification yielded two separable clusters. For pixels with high transmission, the contrast between the absorption features was diminished, so that the classification was less precise. The resulting annotated false-color map shown in Fig.~\ref{fig:microplastics}b shows that each particle was recognized unambiguously as one polymer type. This simple analytical task demonstrated that the hyperspectral imaging setup yields data with sufficient quality to conduct sample annotation and classification based on quantitative criteria. We are therefore convinced that more sophisticated methods of component analysis and annotation procedures can also be performed with our hyperspectral imaging technique.

\subsection{Rat liver tissue}

\begin{figure}[!htbp]
\centering
\includegraphics[width=13cm]{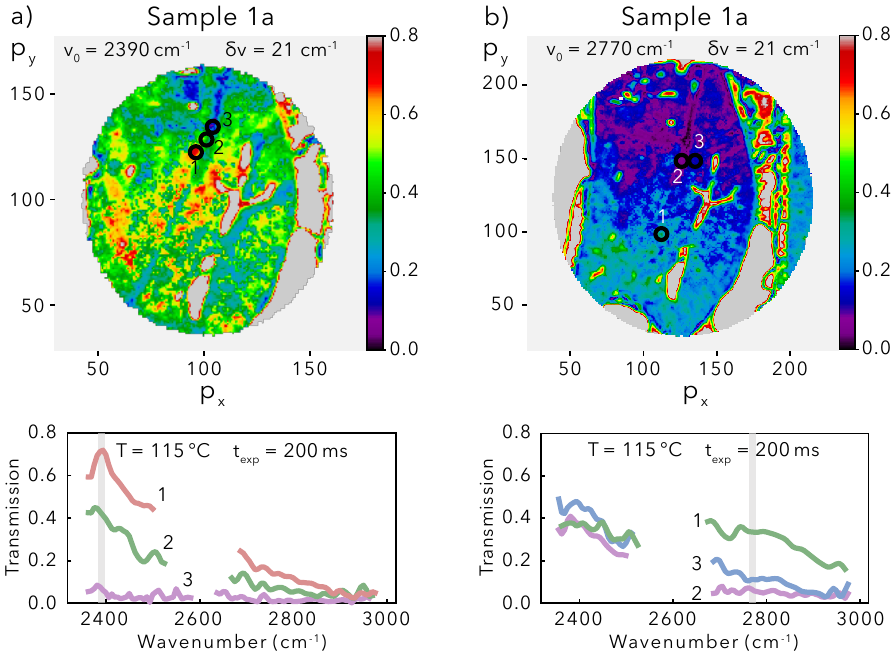} 
\caption{\small{{\bf Hyperspectral analysis of rat liver tissue:} a) Spatial clusters of high transmission were found in contrast-rich spectral images around 2390 cm$^{-1}$ (pixel positions are indicated by black circles and the single pixel spectra correspondence is given by the numbering). These clusters were spatially correlated between consecutive slices but uncorrelated for nonconsecutive slices. By comparison, the samples featured lower contrast and lower maximum transmission values in the spectral band around 2800 cm$^{-1}$ as indicated by the spectral image in b) and the three single pixel spectra of subfigures a and b, respectively. The stark (spectral transmission) contrast observed for some parts of the tissue was significant beyond the level of (Rayleigh) scattering-induced losses.}} 
\label{fig:rat}
\end{figure}

A much anticipated application case is the label-free hyperspectral acquisition on histological samples, where molecular component-specific absorbances provide mid-IR contrast to tissue that is nearly monochrome at visible and near-IR wavelengths. Through the visualization of these contrasts and assisted by reference databases, the distribution of major biomolecule groups such as proteins, lipids, nucleic acids and carbohydrates are selectively addressed and individual components can be classified. Using the spectral mid-IR data, pathological tissue changes such as cancer can be diagnosed and molecular subtypes relevant to prognosis can be determined. 

In our experiment, the hyperspectral acquisition was performed on histological samples of rat liver mounted on the idler-arm protected-silver scanning mirror. Figure \ref{fig:rat} gives two wavenumber-specific transmission images together with a set of pixel-specific transmission spectra. Through analysis of the latter, features of high transmission around 2390 cm$^{-1}$ were identified and distinguished from the tissue's increased overall opacity but reduced wavenumber-dependency in the spectral transmission around 2770 cm$^{-1}$. In the spectral image in Fig.~\ref{fig:rat}a, these spectral features appear to be spatially clustered as islands with red to yellow coloring that predominantly cover the central region of the field of view. By contrast, no comparable transmission levels were found in the spectral image in Fig.~\ref{fig:rat}b that shows sample 1a plotted around 2770 cm$^{-1}$. Due to the lack of a bio-tissue related reference database, the chemical composition of the so identified high transmission clusters could not straightforwardly be annotated. However, the steep transmission decrease between 2390 and 2770 cm$^{-1}$ beyond a $\lambda^{-2}$-proportionality - as it would be expected for losses that merely stemmed from Rayleigh scattering - support the interpretation of these hyperspectral features as expressions of composition-specific absorption. 

The significantly higher opacity in the shortwave band around 2770 cm$^{-1}$ might have resulted from a multitude of hydrocarbon-related absorption features, while the variation to their longwave tail might have produced the stark (transmission) contrast around 2390 cm$^{-1}$; suggesting that lipid deposition was responsible for these features. Furthermore, comparisons of consecutive with nonconsecutive tissue slices confirmed the spatially clustered feature of high transmission around 2390 cm$^{-1}$ throughout multiple samples, see Supplemental Information for additional plots. Here, a strong spatial correlation of the spectral features was found only in consecutive slices with significant differences between farther separated layers. In line, lipid storage in the liver shows regional differences even in healthy tissue \cite{doi:10.1126/sciadv.abq2937} as well as considerable heterogeneity across hepatocytes \cite{HERMS20131489}. Taken together, these findings   further indicated that the observed spectral features genuinely stemmed from the tissue as opposed to arising from measurement artifacts or resulting merely from stochastic parameters of the preparation routine.

\section{Conclusion}
Through the combination of quantum imaging with undetected photons and quantum Fourier transform infrared spectroscopy, we have realized a novel mid-IR hyperspectral acquisition method. For the experimental design, the spectral properties and imaging capability of the nonlinear emission could be modeled precisely and adjusted according to the requirements of the specific samples. Due to the inherently low-noise generation of spontaneous parametric down conversion, the acquisition technique yielded near-ideal performance close to the shot noise limit of the spectroscopic illumination-source. Therefore, the measurements' squareroot-like SNR-scaling holds great potential for accelerating the overall acquisition through sacrificing sensitivity and enhancing the pump power. Moreover, the robustness of the experimental technique was demonstrated through its application to the hyperspectral analyses of polymer, microplastics and organic tissue samples. Here, close coincidence with data from commercial-FTIR reference measurements of the same samples was found for bulk polymer films. Microplastics particles of different polymer species could then be annotated on the basis of the spectral fingerprints obtained from the reference data. Finally, the identification of reproducible hyperspectral features in biological tissue demonstrated the  method's suitability for highly scattering media. Given the limited spatial resolution of the optical arrangement, individual components around the hydrocarbon vibrational modes could not be isolated and distinguished. Nevertheless, spectral features with high transmission around 2400 cm$^{-1}$ also appeared spatially clustered within the tissue, in line with the known heterogeneity of lipid storage in liver tissue. Moreover, the clustering showed high/low spatial correlations for consecutive/nonconsecutive tissue slices. Since the observed transmission contrast exceeded the scattering loss-related wavelength scaling throughout the probed spectral range, our analysis revealed genuine hyperspectral features of the organic tissue and thereby served the anticipated purpose of providing a label-free mid-IR spectral imaging technique for photosensitive samples, while operating near the shot noise limit of the illumination source.

\end{doublespacing}

\renewcommand\refname{References}
\begin{footnotesize}
\bibliographystyle{unsrt.bst} 
\textnormal{\bibliography{localbibliography.bib}}
\end{footnotesize}
\newpage
\section{Supplemental Information}
\subsection{Nonlinear emission: Image resolvability and (temperature-tunable) spectral coverage}

\begin{figure}[!htbp]
    \centering
\includegraphics[width=15cm]{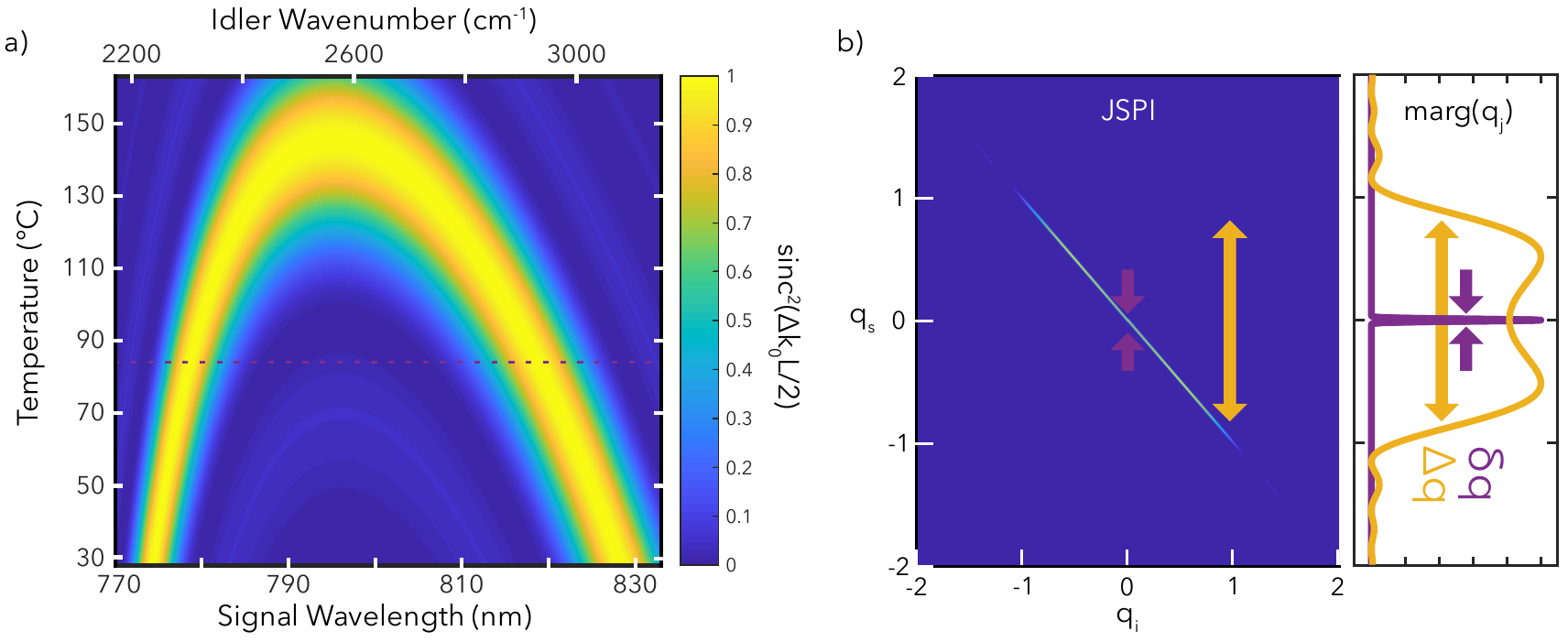}
   \caption{\small{{\bf Nonlinear emission modeling:} a) For a poling period of $\Lambda_0 =$ 20.2 \textmu m, highly broadband and non-degenerate phasematching is satisfied with idler tunability from 2200 to 3200 cm$^{-1}$ within a temperature range of 30 to 130 $^{\circ}$C. Here, a pump wavelength of $\lambda_{p0} =$ 660 nm enables group velocity matched dispersion between the mid-IR idler and the near-IR signal. b) The imaging capability is calculated from the phase matching bandwidth ($\Delta q$) and the conditional momentum uncertainty ($\delta q$) of the joint spatial intensity (JSPI) that determine the field of view and the spatial resolution in the far-field of the crystal. From the ratio of these two quantities, 80 spatial modes are expected in each transverse direction of the image.}}
    \label{fig:dispersionmodel}
\end{figure}

Taking into account precise dispersion data of the KTP crystal, collinear phasematching can be satisfied for the desired combination of pump, signal and idler wavelengths ($\lambda_{j0}$), when the poling period $\Lambda_0$ is chosen according to
\begin{equation}
    \frac{\Delta k_0}{2\pi} = \frac{n(\lambda_{p0}, T)}{\lambda_{p0}} - \frac{n(\lambda_{s0}, T)}{\lambda_{s0}} - \frac{n(\lambda_{i0}, T)}{\lambda_{i0}} - \frac{1}{\Lambda_0}  = 0 \quad . \label{eq1}
\end{equation}
In our case, the condition for group velocity matched signal-idler dispersion demands a practical choice for the pump wavelength of $\lambda_{p0} \approx$ 660 nm. Choosing a corresponding poling period of $\Lambda_0 =$ 20.2  \textmu m, then yields an idler tunability between ambient CO$_2$ ($\nu_i \lesssim$ 2400 cm$^{-1}$) and hydrocarbon-related absorption bands (2800 to 3100 cm$^{-1}$) for an easily accessible temperature range (30 to 130 $^{\circ}$C). Figure \ref{fig:dispersionmodel}a gives signal and idler power spectral densities (PSD) that follow from equation \ref{eq1} when energy conservation is imposed through $\lambda_{p0} = (\lambda_{s0}^{-1} + \lambda_{i0}^{-1})^{-1}$ as functions of the crystal temperature $T$. 
For spectral tunability of the nonlinear emission, the crystal is mounted in a PID-controlled oven with the operation temperature set to 87 degrees Celsius (115 $^{\circ}$C) to optimally address the spectral features of different polymer (bio-tissue) samples.
Modeling the vectorial properties of the generated pair photons for the given crystal with ($\Lambda = \Lambda_0$ and $T=T_{\textrm{exp}}$) the imaging capability in the Fourier plane of the crystal can then be qualified from the joint spatial intensity (JSPI):
\begin{equation}
    JSPI = \left| \textrm{sinc}( [ \frac{\left|q_s\right|^2}{2 k_{s0}} + \frac{\left|q_i\right|^2}{2 k_{i0}} - \frac{\left|q_s + q_i\right|^2}{2 k_{p0}} + \Delta k_0] \cdot L_{\textrm{c}}/2) \cdot \Phi_p(q_i+q_s) \right|^2
\end{equation}
Here, the phasematching function $\Psi = \textrm{sinc}(\Delta k(q_s, q_i)L_{\textrm{c}}/$2) that ensures momentum conservation between the three nonlinearly coupled frequencies is expressed as a function of their transverse wavevector components $q_j$. Analogous to the energy conservation for the spectral densities, the transverse momentum (uncertainty) associated with the pump waist size imposes a restriction on the sum of the signal and idler JSPI variables, namely:
\begin{equation}
\Phi_p =    \exp(-(\left| q_s+q_i \right|^2)\cdot \omega_p^2/4) \quad.
\end{equation}
Finally, in the Fourier plane, the imaging capability (resolvability) is determined by the ratio of phase matching bandwidth ($\Delta q \propto L_{\textrm{c}}^{-1/2}$) and the transverse momentum uncertainty scaled by the pump function waist ($\delta q  \propto \omega_p^{-1}$). The imaging performance in terms of the spatial resolution and field of view (FOV) then arises through the propagation of the transverse momenta into the far-field of the crystal considering the employed focusing optical elements that translate these momenta to positions within the Fourier plane.

\subsection{Classification criterion for polystyrene and polypropylene samples}

\begin{figure}[!htbp]
    \centering
    \includegraphics[width=14cm]{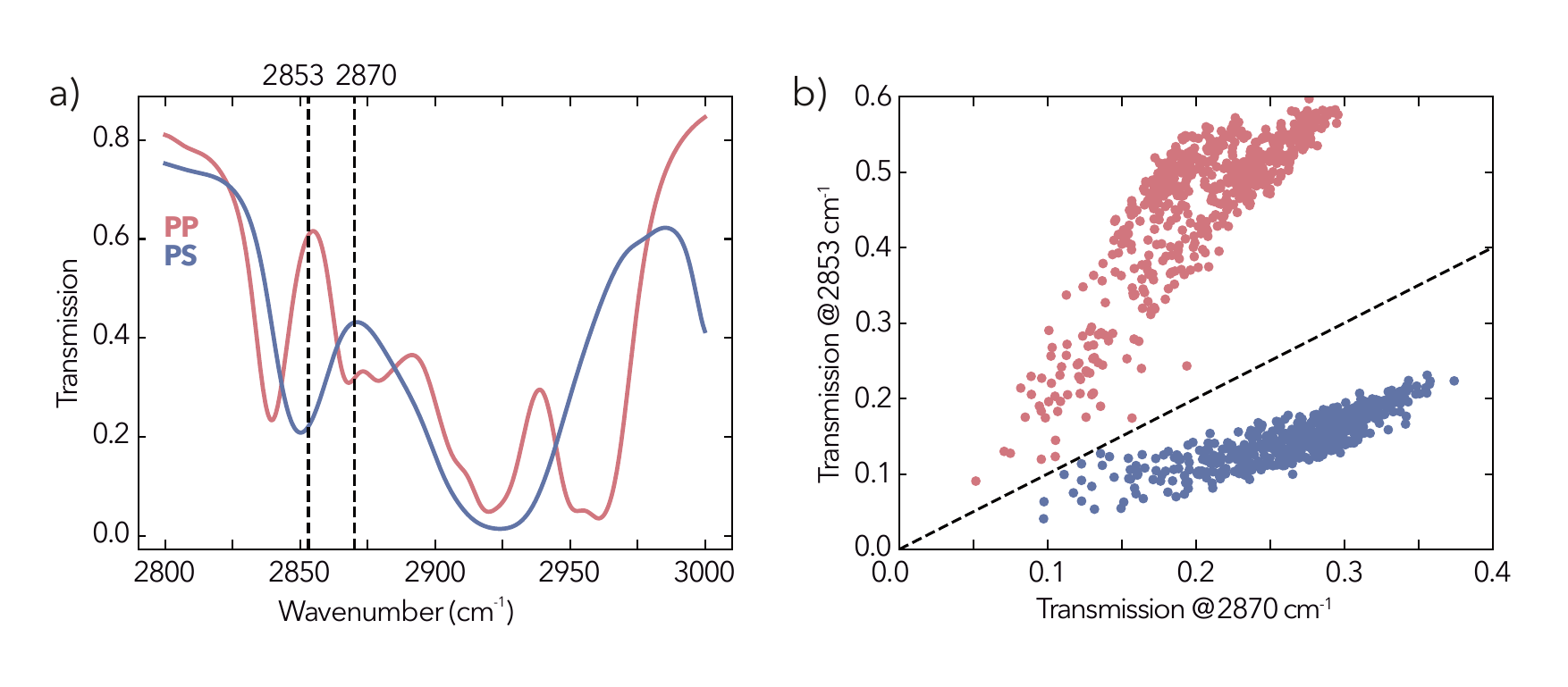}
    \caption{\small{{\bf Simple classification scheme to distinguish polystyrene (PS) and polypropylene (PP):} a) FTIR reference transmission spectra of the sample polymer films show a inverse relation of the spectral transmission at 2855 and 2870 cm$^{-1}$, respectively. b) Experimental hyperspectral imaging data of the corresponding samples revealed a clustered distribution in the joint spectral transmission diagram at these two wavenumbers and a linear decision boundary was found.}}
    \label{fig:classification2}
\end{figure}
For the two-species microplastics sample, the hyperspectral data was used to distinguish polystyrene- from polypropylene-like absorption signatures for each pixel with a significant transparency reduction (to below 80\%). The applied distinction criterion was first defined and evaluated with regard to the bulk polymer films for which both hyperspectral imaging data with medium spectral resolution as well as FTIR reference data had already been recorded.
In the reference spectra shown in Fig.~\ref{fig:classification2}a, the polystyrene (PS, blue curve) showed an absorption dip at 2850 cm$^{-1}$ and a broad absorption feature around 2900 to 2950 cm$^{-1}$, while polypropylene (PP, red curve) featured a narrow absorption  around 2839 cm$^{-1}$, higher transmission around 2855 cm$^{-1}$ and another pronounced absorption structure around 2867 to 2878 cm$^{-1}$. These signatures were used to distinguish PP and PS in a simplified scheme: we classified each pixel as PS-containing if $T$(2853 cm$^{-1}$) < $T$(2870 cm$^{-1}$) was satisfied and otherwise as PP-containing.

As a preliminary test of this distinction criterion, the hyperspectral data of the four bulk polymer film sample was evaluated in this manner, since in this case the pixel-to-species correspondence was known. Figure~\ref{fig:classification2}b shows the joint spectral transmission values of each PS-containing pixel in blue and each PP-containing pixel in red. Each PP-like pixel was found to appear in the upper left segment as opposed to the PS clustering in the lower right section allowing for a linear decision boundary with a slope of one in the joint (two-point) transmission plot. This finding confirmed that the two-point spectral transmission comparison was a suitable classification criterion to distinguish PP and PS for pixel-based data; independent of the absolute transmission level, which showed slight deviations between FTIR reference measurements and the hyperspectral acquisition due to scattering and polarization effects. For highly opaque pixels, the distinction became less sharp, as the contrast between the absorption features decreased such that both species clustered closer to the decision boundary. Given the successful classification of the bulk polymer films, the scheme was subsequently applied to the microplastics sample.

\subsection{Extended data on rat liver tissue sample}

\begin{figure}[!htbp]
\centering
\includegraphics[width=13cm]{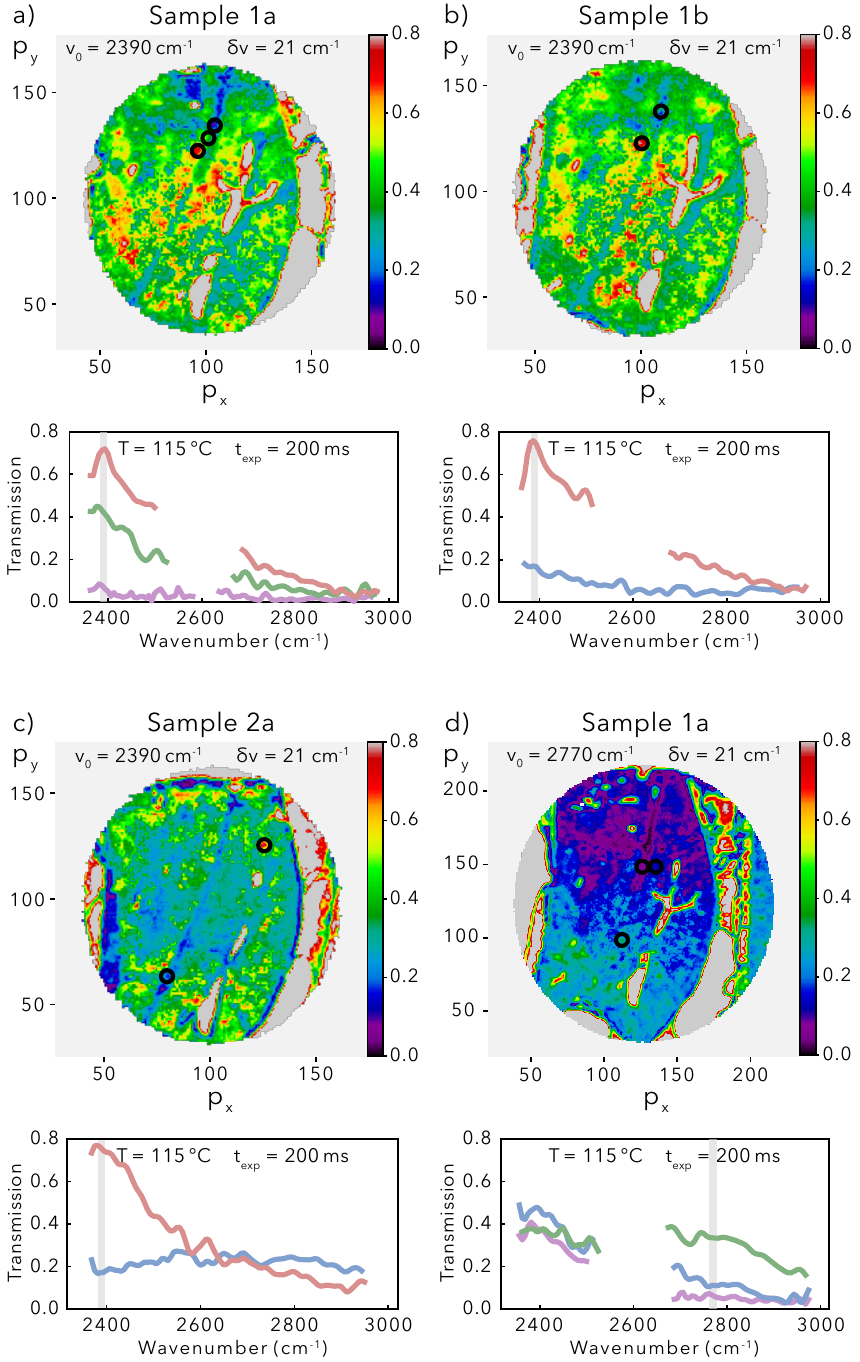} 
\caption{\small{{\bf Hyperspectral analysis of rat liver tissue:} a) Spatial clusters of high transmission were found in contrast-rich spectral images around 2390 cm$^{-1}$ (pixel positions are indicated by black circles and the single pixel spectra correspondence is given by the resembling color of the plotted solid line and the encircled pixel). These clusters were spatially correlated between consecutive slices (samples 1a \& 1b) but uncorrelated for nonconsecutive slices (sample 1 \& 2). By comparison, the samples featured overall lower transmission and lower contrast in the spectral band around 2800 cm$^{-1}$ as indicated by the spectral image in subfigure d and the three single pixel spectra of subfigures a and d, respectively. The stark (transmission) contrast was significant beyond the level of Rayleigh scattering induced losses. Altogether with the high degree of spatio-spectral correlations between consecutive slices, these observations supported the interpretation of the hyperspectral contrast to genuinely be related to the composition of the tissue.}} 
\label{fig:rat2}
\end{figure}

In our experiment, the hyperspectral acquisition was performed on histological samples of rat liver mounted on the idler-arm protected-silver scanning mirror. Figure \ref{fig:rat2} gives a series of wavenumber-specific transmission images together with pixel-specific transmission spectra. Through analysis of the latter, features of high transmission around 2390 cm$^{-1}$ were identified and distinguished from the tissue's increased overall opacity but reduced wavenumber-dependency in the spectral transmission around 2770 cm$^{-1}$. In the spectral images of subfigures \ref{fig:rat2}a to c, these spectral features appear to be spatially clustered as islands with red to yellow coloring that predominantly cover the central region of the field of view for sample 1a and 1b and the outer FOV regions for sample 2, respectively. By contrast, no comparable transmission levels were found in the spectral image of subfigure d that shows sample 1a plotted around 2770 cm$^{-1}$. Due to the lack of a bio-tissue related reference database, the chemical composition of the so identified high transmission clusters could not straightforwardly be annotated. However, the steep transmission decrease between 2390 and 2770 cm$^{-1}$ beyond a $\lambda^{-2}$-proportionality - as it would be expected for losses that merely stemmed from Rayleigh scattering - support the interpretation of these hyperspectral features as genuine expressions of composition-specific absorption. The significantly higher opacity in the shortwave band around 2770 cm$^{-1}$ might have resulted from a multitude of hydrocarbon-related absorption features, while the variation to their longwave tail might have produced the stark (transmission) contrast around 2390 cm$^{-1}$; suggesting that lipid deposition was responsible for these features. Furthermore, comparisons of consecutive (sample 1a \& 1b) with nonconsecutive (sample 2a) tissue slices confirmed the spatially clustered feature of high transmission around 2390 cm$^{-1}$ throughout multiple samples. Here, a strong spatial correlation of the spectral features was found only in consecutive slices with significant differences between farther separated layers. In line, lipid storage in the liver shows regional differences even in healthy tissue as well as considerable heterogeneity across hepatocytes. Taken together, this finding further indicated that the observed spectral features genuinely stemmed from the tissue as opposed to arising from measurement artifacts or resulting merely from stochastic parameters of the preparation routine.

\subsection{Biological tissue preparation}
In accordance with the 3R principle, we used tissue specimens from a recently published animal experiment. All experiments were performed in accordance with the German/ European law for animal protection and approved by the local ethics committee (LaGeSo, G0019/21). In a 3R approach, we used liver tissue not needed in the published animal experiment. In short, 7-week-old Sprague–Dawley rats, overexpressing human angiotensinogen and renin, were perfused with physiological saline solution in isoflurane narcosis (2\% to 2.5\%) and analgesia with 0.05 mg/kg bodyweight buprenorphine via the left ventricle. Liver tissue was excised and fixed for 6 to 18 hrs in 4\% paraformaldehyde (PFA, Science Services) in phosphate-buffered saline (PBS, Thermo Fisher) at 4 $^{\circ}$C. After washing in PBS, tissue was cryoprotected using a sucrose gradient (15\% 8 to 12 h, 30\% 12 to 18 h) and frozen in Tissue-Tek O.C.T. (Sakura Finetek) using -40 $^{\circ}$C 2-methylbutan (Merck Millipore).
Tissues were cut into 7 \textmu m cryosections at -20 $^{\circ}$C using a Leica CM3050 S Cryostat. Cryosections were prepared directly on mirrors. Cryocuts were subsequently stored at -20 $^{\circ}$C until further use.
\end{document}